\documentstyle[12pt,preprint]{aastex}

\begin{document}

\title{Ultraluminous x-ray sources, high redshift QSOs and active galaxies}

\author{G. Burbidge\altaffilmark{1}, E.M. Burbidge\altaffilmark{1}, H.C. Arp\altaffilmark{2},
\& W.M. Napier\altaffilmark{3}}

\altaffiltext{1}{Center for Astrophysics and Space Sciences 0424 University of California San Diego, CA 92093-0424, USA}
\altaffiltext{2}{Max-Planck-Institut für Astrophysik 85741 Garching, Germany}
\altaffiltext{3}{Centre for Astrobiology, Cardiff University, 2 North Road,
Cardiff CF10 3DY, Wales, UK}

\begin{abstract}

It is shown that all of the 32 point X-ray sources which lie within about 10$^\prime$ of
the centre of nearby galaxies, and which have so far been optically identified, are high
redshift objects -- AGN or QSOs. Thus the surface density of these QSOs $\rho \simeq 0.1$
per square arc minute.  Some of them were originally discovered as X-ray sources and
classified as ultraluminous X-ray sources (ULXs), nearly all of which lie near the
centers of active galaxies. We demonstrate that this concentration around galactic nuclei
is of high statistical significance: the probability p that they are accidental lies in
the range 10$^{-3}-10^{-4}$, and apparently this excess cannot be accounted for by
microlensing.
\end{abstract}


\section{Introduction and data}

Optical identifications of compact X-ray sources have shown that they may be (a) binary
star systems where accretion of matter from the optically luminous star on to a companion
neutron star or black hole is the primary energy source, (b) quasi-stellar objects with a
wide range of redshifts, or (c) the nuclei of active galaxies. High resolution studies
made with Einstein, Rosat, Chandra and XMM have shown that powerful point X-ray sources
displaced from the nuclei of the galaxies can be found apparently very close to or in the
main bodies of many comparatively nearby galaxies (cf Colbert et al, 2002; Fabbiano et
al, 2001; Foschini et al, 2002a\&b; Kaaret, Makishima, et al, 2000; Strickland et al,
2001; Wu et al 2002; Zezas et al, 2002).  The distances of all these galaxies are so
large that the luminosities of these X-ray sources must be $\simeq10^{39}~erg\,s^{-1}$.
Thus if they are binaries similar to those identified in our own galaxy they may be black
hole accretion sources with masses in the range 10$^3$--10$^4$\,M$_\odot$, or possibly
accretion sources with lower masses but which are beamed. While there have been many
x-ray studies of these sources, very few have been optically identified.

In 2003, we suggested that some of them might, instead, be X-ray
emitting QSOs with a wide range of redshifts (Burbidge et al 2003).
In that case they might have been identified first as QSOs very
close to the nuclei of galaxies, and then found later to be X-ray
sources. Alternatively, from the X-ray detections of ULX sources,
accurate enough positions might enable us to identify optical
objects which turn out to be QSOs.

It turns out that both methods have worked. We have surveyed the literature and show in
Table~1 a list of all the known X-ray emitting QSOs which lie within $\sim$10$^\prime$ of
the centers of comparatively nearby galaxies. From the redshifts of these galaxies and
using a Hubble constant of 60~km~s$^{-1}$\,Mpc$^{-1}$ (Tamman et al 2002), it can be seen
that all of the galaxies lie within about 35\,Mpc. Nearly all of them show signs of
activity in their nuclei.

All types of galaxies have been surveyed for ULXs and there is some indication that more
ULX's are found in elliptical galaxies than in spirals (cf Colbert \& Ptak 2002). Thus it
may well be that a significant number of the ULXs are black hole accretion binaries since
there is no evidence for nuclear activity in most of those galaxies. So, as far as we are
aware there are, as yet, no optical identifications of any ULXs which are binary systems.
However the brightest ULX in M82 has been shown from X-ray observations to show
variations giving an orbital period of about 62 days (Kaaret, Simet and Lang 2006).  Of
course this is not an optical identification but it does suggest that this is a binary
source, and we do not doubt that many ULXs still to be identified are massive binaries.

However, the main point of this paper remains.  Table 1 demonstrates that many ULX
sources are high redshift QSOs as we suggested in 2003 (Burbidge et al 2003)

\section{Concentrations of ULX QSOs around nearby galaxies }

There have been numerous claims in the literature that QSOs tend to concentrate around
nearby, active galaxies (cf Radecke and Arp, 1997). Since \emph{only} ULX QSOs within 10
arc minutes of a galactic nucleus were selected in the present study, the hypothesis that
they are significantly concentrated around galaxies should be tested by comparison with
other samples of X-ray emitting QSOs. An ideal 'null sample' would be the general
background of ULX QSOs. Instead, comparison is here made with the background density of
QSOs in general since this appears to be well-known as a function of magnitude (Kilkenny
et 1997; Myers et al. 2005). Since only a fraction of background QSOs are X-ray sources,
the statistical significance of any concentration will thereby be underestimated.

In Fig.~1 the magnitude distribution of the Table~1 ULX QSOs is compared with that
derived from the Sloan Digital Sky Survey (Myers et al. 2005). It is clear that the
sample contains a disproportionately large number of very bright QSOs.  A t-test reveals
the excess to be significant at a confidence level $\sim10^4$ to one. The median
magnitude from Table~1 is 19.3, although absorption by the intervening galaxies may imply
that some or all are intrinsically brighter. The total number of galaxies which have been
surveyed using X-ray telescopes to find ULXs is about 200. However, very little work has
been done on optical identifications in general, and all of the positively identified
ULXs are those in Table~1. If we conservatively assume that 20 galaxies have been
surveyed in arriving at these identifications, then the area of sky searched for Table~1
is about 1.75 square degrees. Observer bias could have entered through selective
surveying of ultraluminous sources close to galactic nuclei. Thus in NGC~720, four QSOs
have been identified within 4$^\prime$ of the nucleus but none are listed between
4$^\prime$ and 10$^\prime$. Does this reflect a real concentration in the sky, or a
choice of targets by the observer? The question can be circumvented by testing any
apparent concentration against the total background count expected for any prescribed
limiting magnitude, obtained from homogeneous QSO surveys.

There is clearly a significant surfeit of very bright QSOs and of QSOs very close to the
galactic nuclei. The former result is illustrated in Fig.~2 for QSOs with $b<$17.5, which
reveals an excess of such objects at all angular distances out to 10$^\prime$. There are
two such objects within 3$^\prime$ as against an expectation of 0.027 (probablility
$p\sim 3.6 \times 10^{-4}$) increasing to 4 within 10$^\prime$ as against an expectation
of 0.297 ($p\sim 2.6 \times 10^{-4}$). Also, there are five QSOs brighter than 19.0
within 3$^\prime$ of the nuclei as against an expectation of 0.67, a result which has
$p\sim 6.2 \times 10^{-4}$.
These probabilities are likely to be conservative in that: \\
(i) they are derived assuming 100\% discovery of bright ULX QSOs out to 10$^\prime$; \\
(ii) comparison is made with the total QSO background density rather
than the background of ultraluminous QSOs; and \\
(iii) in the null hypothesis no correction for dimming of the QSOs
is made although on the conventional interpretation they are seen
through the galaxies ($\la$100~kpc).

The referee has made the point that we have assumed here that only about 20 out of
$\simeq200$ spiral galaxies are know to contain ULX sources which are high redshift QSOs.
He is correct in pointing out that if all 200 had been studied, the total area used in
our calculation would be 17.5 square degrees.  Thus if no more QSOs were found in the
remaining galaxies the probabilities that these concentrations are accidental would be
increased by factors of 10.  However, we strongly suspect that many more QSOs will be
found in the galaxies still to be surveyed. But, very conservatively we can conclude that
the density $\rho$ of bright QSOs in the nuclear regions of galaxies lies somewhere in
the range $0.1>\rho>0.01$ per square arc minutes.

\section{Discussion and Conclusion}

Assuming the halo of each galaxy to have a mass 10$^{12}$\,M$_\odot$
in the form of microlenses, then the 14 galaxies have an aggregate
microlensing area equivalent to a single object of Einstein radius
$\Theta_E\sim$85~arcsec, corresponding to an area $\sim$0.024~sq~deg
or about 2 percent of the area under consideration.

To boost the expected number of bright QSOs (say $<$17.5~mag, with background density
$\sim$0.15 per square degree) from the expectation value $\sim$0.3 to the observed 4
would require the lensing of faint QSOs within the Einstein radii with background
densities at least two orders of magnitude higher, i.e. magnitude 20.5 or fainter.
Microlensing both enhances the brightness of faint background QSOs and lowers their
density over the sky.  Because of the shallowness of the slope at the faint end of the
QSO distribution -- below the knee at $b_0 =19.1$ -- microlensing creates a deficit
rather than an excess of counts (Myers et al. 2005).

Myers et al (2005) have remarked that standard $\Lambda$CDM models have difficulty
producing the QSO excess they find on 100~kpc scales, and remark that either bias is
strongly scale dependent, or``there exists an unexpected, strong systematic effect
inducing positive correlations between QSOs and foreground galaxies". The above argument
suggests that even with halo masses $\sim 10^{12}$\,M$_\odot$, microlensing is unable to
account for the observed excess. This would seem to support the view that at least a
proportion of the ULX QSOs are physically close to the nuclei of galaxies.  Clearly some
of ULX sources are high redshift QSOs closely associated with the galaxies.

\clearpage

\begin{figure}
\begin{minipage}[b]{1.0\linewidth}
\center{\includegraphics[angle=0,width=1.0\linewidth]{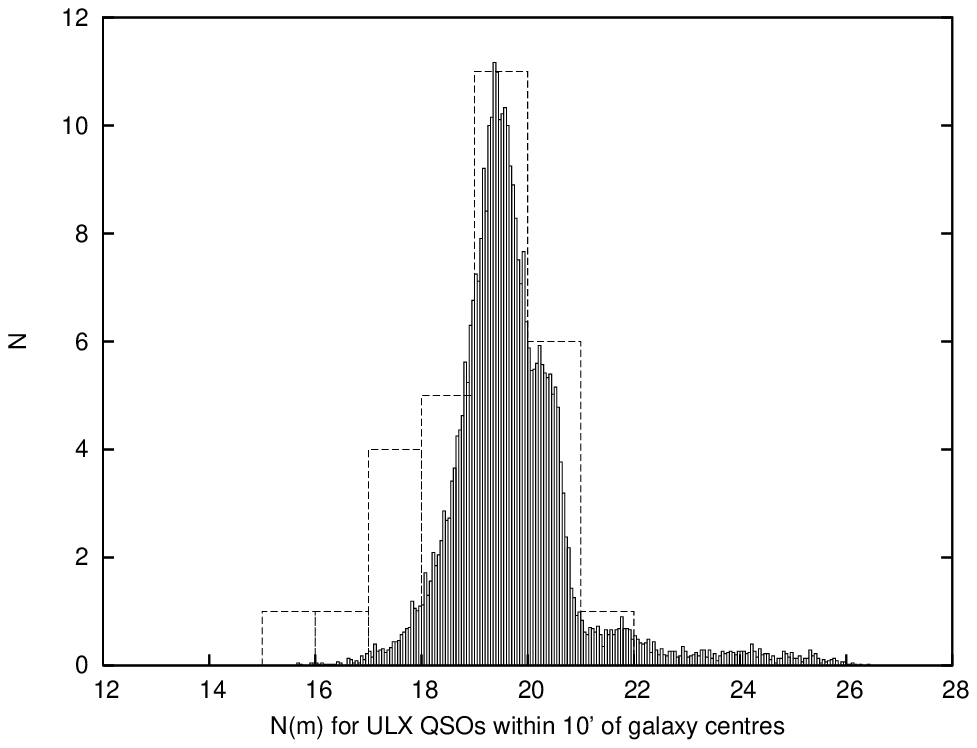}}
\end{minipage}
\caption{Magnitude distribution of ULX QSOs in Table~1 (histogram)
as compared with that from the Sloan Survey, suitably normalised
(solid curve).}
\end{figure}

\clearpage

\begin{figure}
\begin{minipage}[b]{1.0\linewidth}
\center{\includegraphics[angle=0,width=1.0\linewidth]{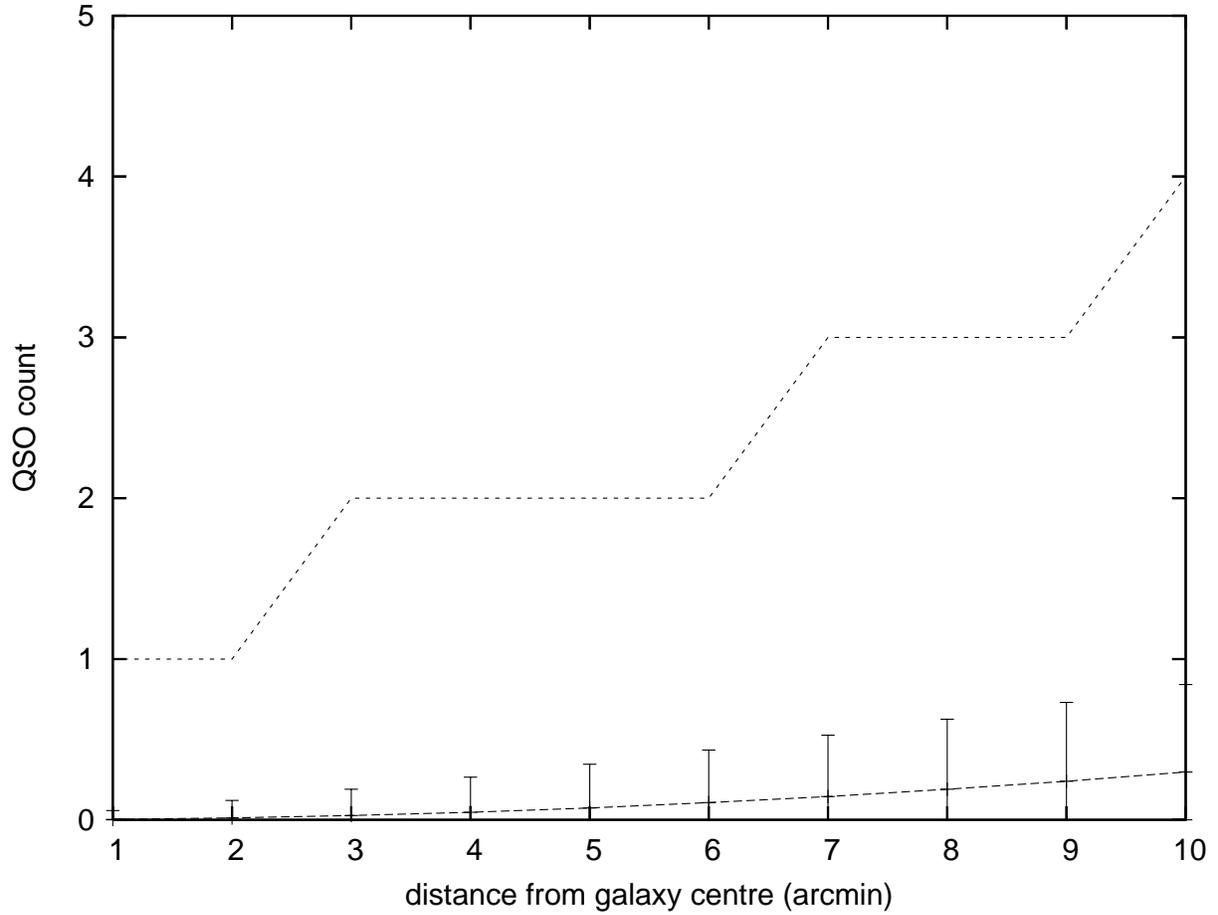}}
\end{minipage}
\caption{Upper curve: cumulative numbers of ULX QSOs with $b<$17.5
as a function of angular distance from the galaxy nuclei. Lower
curve: expected background count for all QSOs, whether ULX or not,
assuming 100\% completeness of discovery. Error bars are 1$\sigma$
limits based on Poissonian estimates.}
\end{figure}

\clearpage

\begin{table*}
\caption{X-Ray Emitting QSOs Lying Within 10$^\prime$ of Galaxy
Centers}
\begin{tabular}{llrrrrrl}
GALAXY   & TYPE        & $z_G$  & $z_Q$& $m_Q$ & $\Theta$ & $d$ (kpc) & SOURCE \\
\hline

NGC 720  & E4          & 0.0058 & 0.39  & 21.5  &  1.9 & 15.8 &
Burbidge et al (2005b) \\

         &             &        & 0.39  & 22.0  &  2.3 & 19.6 \\

         &             &        & 0.96  & 19.2  &  2.7 & 27.9 & Arp et
al (2004) \\

         &             &        & 2.22  & 20.6  &  3.4 & 35.1 \\ *[5pt]

NGC 1073 &  SB(rs)c    & 0.0040 & 1.41  &  20.0 &  1.4 & 12.3 & Arp
\& Sulentic (1979) \\

         &             &        & 0.56  &  18.8 &  2.1 & 18.4 &
Burbidge et al (1979) \\ *[5pt]

NGC 1365 &(R')SBb(c)b  & 0.0055 & 0.90  &  19.7 &  7.7 & 95.9 &
LaFranca et al (2002) \\

         &    Sy1.8    &        & 0.31  &  18.0 & 12.4 & 154.5 \\ *[5pt]

NGC 3079 &  SB(S)c     & 0.0037 & 1.04  &  19.1 &  8.8 & 94.7 &
Arp et al (2005) \\

         &    Sy2      &        & 0.680 &  21.0 &  7.5 & 80.7 &    Burbidge et al (2005a) \\

         &             &        & 0.673 &  18.5 & 10.0 &  107.6 \\

         &             &        & 0.216 &  17.5 & 13.0 &   140.0
 \\*[5pt]

NGC 3628 &  Sab pec    & 0.0028 & 0.995 &  20.1 &  3.0  & 18.3 &
Arp et al (2002) \\

         &    Liner    &        & 2.150 &  19.5 &  4.2  & 25.7 \\

         &             &        & 0.981 &  19.2 &  5.5 & 33.6 \\

         &             &        & 0.408 &  19.6 &  5.6 & 34.2 \\

         &             &        & 2.430 &  19.9 &  6.0 & 36.7 \\

         &             &        & 2.060 &  19.6 &  7.9 & 48.3 \\

         &             &        & 1.940 &  18.3 & 12.3 & 75.2 \\ *[5pt]

NGC 4039 &SA(S')m Liner& 0.0055 & 0.26  &       & 0.34 &   3.4 & Clark et al (2005) \\

NGC 4151 & (2')SAB(rs) & 0.0033 & 0.613 &  20.3 &  4.9 & 36.8  &
Page et al (2001) \\

         &    Sy 1.5   &        & 0.022 &  17.2 &  6.8 & 56.1  &  Rector et al
(2000) \\ *[5pt]

NGC 4168 & E2 Sy 1.9   & 0.0074 & 0.217 & 18.7  & 0.75 &   4.8 &
Masetti et al (2003) \\

NGC 4203 & SABO Liner  & 0.0036 & 0.614 & 17.5  &  2.1 & 13.1  &
Knezk \& Bregman (1998) \\

NGC 4258 &  SAB(s)bc   & 0.0015 & 0.398 &  20.4 &  8.6 & 39.6  &
Burbidge (1995) \\

         &   Sy 1.9    &        & 0.653 &  19.9 &  9.7 & 44.7  & Mironi (2003) \\

         &             &        & 0.520 &  17.0 &  9.7 & 44.7 \\ *[5pt]

NGC 4319 &SB(r)ab AGN  & 0.0045 & 0.071 &  15.2 &  0.7 & 4.6   &
Stocke et al (1991) \\

NGC 4374 & E1 Sy2      & 0.0035 & 1.25  &  18.5 &  2.4 & 15.3  &
Burbidge et al (1990) \\

NGC 4698 & SA(S)ab Sy2 & 0.0033 & 0.43  &  20.5 &   1.2 & 5.7  &
Foschini et al (2002) \\

NGC 7319 & SB(s)bc Sy2 & 0.0023 & 2.11  &  21.8 &  0.13 & 4.3  &
Galianni et al (2005) \\

\end{tabular}
\end{table*}


\begin{thebibliography}{}

\bibitem{}
{Arp, H.C., Gutierrez, C.M. \& Lopez-Corredoira, M., 2004, A\&A,
418, 877}

\bibitem{}
{Arp, H.C. et al 2002, A\&A, 391, 833}

\bibitem{}
{Arp., H.C. \& Sulentic, J., 1979, ApJ., 229, 496}

\bibitem{}
{Burbidge, E.M. 1995, A\&A, 2 95, L1}

\bibitem{}
{Burbidge, E.M., Junkkarinen, V., Koski, A., 1979, ApJ, 233, L97}

\bibitem{}
{Burbidge, E.M., Burbidge G., Arp, H.C., \& Napier, W., 2005a,
astro-ph/0510815}

\bibitem{}
{Burbidge, E.M., Arp, H.C., \& Gutierrez, C., 2005 (in preparation)}

\bibitem{}
{Burbidge, G., Burbidge, E.M., \& Arp, H.C., 2003, A\&A, 400, L17}

\bibitem{}
{Clark, D.M. et al 2005, ApJLetter 631, 100}

\bibitem{}
{Colbert, E., \& Ptak, A., 2002, ApJS, 143, 25}

\bibitem{}
{Fabbiano, G., Zezas, A., \& Murray, S.S., 2001, ApJ, 554, 1035}

\bibitem{}
{Foschini, L., Di Cocco, G., Ho, L. C., et al 2002a, arXiv;
[astro-ph/0206418]}

\bibitem{}
{Foschini, L., Di Cocco, G., Ho, L. C., et al 2002b, arXiv;
[astro-ph/0209298]}

\bibitem{}
{Galianni, P., Burbidge, E.M., Arp, H., Junkkarinen, V., Burbidge,
G. et al, 2005 ApJ, 620, 88-94}

\bibitem{}
{Jeltema, E.T., Canizares, C., Buote, D. and Garmire, G., 2003, ApJ,
585, 756}

\bibitem{}
{Kaaret, P., Prestwich, A., Zezas, A., et al. 2001, MNRAS, 321, L29}

\bibitem{}
{Kaaret, P., Simet, M., and Lang, C., 2006, Science, 311, 491}

\bibitem{}
{Kilkenny, D., O'Donoghue, D., Koen, C., et al, 1997, MNRAS, 287,
767}

\bibitem{}
{Knezsk   \& Bregman, J., 1998 La Franca, F., et al, 2002, ApJ.,
570, 100}

\bibitem{}
{Makishima, K., Kubota, A., Mizuno, T., et al, 2000, ApJ, 535, 632}

\bibitem{}
{Masetti, N. et al, 2003, ApJ, 400, L27 Mirani, L., 2003, Ph.D,
Thesis, University of Strasburg}

\bibitem{}
{Myers, A.D., et al, 2005, MNRAS, 359, 741}

\bibitem{}
{Radecke, D., and Arp, H.C., 1997, A\&A, 319, 33}
\bibitem{}
{Strickland, D.K., Colbert, E.J.M., \& Heckman, T.M., 2001, ApJ,
560, 707}

\bibitem{}
{Tammann, G., Rindl, B., Thim, F., Sala, A., and Sandage, A.R., 2002, A New Era in
Cosmology (ed. N. Metcalfe \& T. Shanks), 258}

\bibitem{}
{Wu, H., Xue, S.J., Xia, X.Y., Deng, Z.G., \& Mao, S. 2002, ApJ,
576, 738}

\bibitem{}
{Zezas, A., Fabbiano, G. Rots, A.H., and Murray, S., 2002, ApJ.
Supp. 142, 239}

\bibitem{}
{Zezas, A. \& Fabbiano, G., 2002, ApJ, 577, 726}

\end{thebibliography}
\end{document}